\documentclass[a4paper, superscriptaddress, aps, pre, reprint, showkeys, twoside]{revtex4-1}
\usepackage[utf8]{inputenc}
\usepackage[margin=0.5in]{geometry}
\usepackage{datetime}
\usepackage{amsmath}
\usepackage{amssymb}
\usepackage{enumitem}
\usepackage{graphicx}
\usepackage[all]{nowidow}
\usepackage{hyperref}

\hypersetup{
	colorlinks=true,
    linkcolor=blue,
    urlcolor=black,
    citecolor=blue,
}

\makeatletter
\renewcommand{\p@subsection}{}
\renewcommand{\p@subsubsection}{}
\makeatother

\newcommand{\bs}[1]{\boldsymbol{\mathrm{#1}}}
\DeclareMathOperator{\spn}{span}

\begin{document}

\title{On the efficient numerical computation of covariant Lyapunov vectors}

\author{Jean-Jacq~du~Plessis}
    \affiliation{Nonlinear Dynamics and Chaos Group, Department of Mathematics and Applied Mathematics, University of Cape Town, Rondebosch, 7701, South Africa}
\author{Malcolm~Hillebrand}
    \affiliation{Nonlinear Dynamics and Chaos Group, Department of Mathematics and Applied Mathematics, University of Cape Town, Rondebosch, 7701, South Africa}
\author{Charalampos~Skokos}
    \thanks{Corresponding author: \href{mailto:haris.skokos@uct.ac.za}{haris.skokos@uct.ac.za}}
    \affiliation{Nonlinear Dynamics and Chaos Group, Department of Mathematics and Applied Mathematics, University of Cape Town, Rondebosch, 7701, South Africa}
    \affiliation{Max Planck Institute for the Physics of Complex Systems, Nöthnitzer Str.\ 38, Dresden, 01187, Germany}

\date{\today}

\begin{abstract}
Covariant Lyapunov vectors (CLVs) are useful in multiple applications, but the optimal time windows needed to accurately compute these vectors are yet unclear. To remedy this, we investigate two methods for determining when to safely terminate the forward and backward transient phases of the CLV computation algorithm by Ginelli et al.~\cite{GinelliEtAl2007} when applied to chaotic orbits of conservative Hamiltonian systems. We perform this investigation for two prototypical Hamiltonian systems, namely the well-known H\'enon-Heiles system of two degrees of freedom and a system of three nonlinearly coupled harmonic oscillators having three degrees of freedom, finding very similar results for the two methods and thus recommending the more efficient one. We find that the accuracy of two-dimensional center subspace computations is significantly reduced when the backward evolution stages of the algorithm are performed over long time intervals. We explain this observation  by examining the tangent dynamics of the center subspace wherein CLVs tend to align/anti-align, and we propose an adaptation of the algorithm  that improves the accuracy of such computations over long times by preventing this alignment/anti-alignment of CLVs in the center subspace.
\end{abstract}



\maketitle



\section{Introduction}
Lyapunov exponents (LEs) describe the exponential growth rates of perturbations (the so-called deviation vectors) to a trajectory of a dynamical system in various directions of the tangent space. The existence of these LEs is typically guaranteed by Oseledets' multiplicative ergodic theorem \cite{Oseledets1968}. A so-called covariant Lyapunov vector (CLV) is a unit vector whose growth or decay rate is governed by a single LE when evolved either forward or backward in time. The CLVs are termed ``covariant'' because they evolve consistently with the system’s dynamics: if a set of CLVs is known at a given point along an orbit, their dynamical evolution to another point on the same orbit yields the corresponding CLVs at that new point (up to normalization).  Several efficient algorithms for computing CLVs have been developed in recent decades (see e.g.\ \cite{GinelliEtAl2007,WolfeSamelson2007,KuptsovParlitz2012}). Since the development of such algorithms, CLVs have found a variety of applications, including the study of so-called hydrodynamic Lyapunov modes \cite{YangRadons2010,YangRadons2013,YangRadons2013a}, bifurcation classification \cite{KamiyamaEtAl2014}, geophysical fluids \cite{Frederiksen2023}, and atmospheric models \cite{PazoEtAl2010,HerreraEtAl2011,QuinnEtAl2021}.

In this paper, we propose a simple method for determining when to safely stop the transient phases of the CLV algorithm developed by Ginelli and collaborators (hereafter, the GC algorithm) \cite{GinelliEtAl2007}. As we discuss in Sect.~\ref{sec:theory}, this algorithm includes two transient phases: a forward transient phase where a set of deviation vectors are evolved forward in time until they converge to a particular set of subspaces that form a filtration of the system's tangent space, and a backward transient phase where another set of deviation vectors evolved backward in time converge to the CLVs, which form a splitting of the tangent space. The particular subspaces spanned by these deviation vectors are known to converge to the relevant filtration/splitting subspaces at exponential rates given by the spectral gaps between LEs \cite{GinelliEtAl2013,Noethen2019,Noethen2019a}, but we are unaware of any previous studies that efficiently determine when this convergence has occurred and hence when the transient phases can be terminated without loss of accuracy. Without such an empirical approach, a user of this CLV algorithm might need to guess the appropriate transient lengths to use, which risks wasting CPU time (if over-estimating the transient length) or poorly approximating the CLVs (if under-estimating). Therefore, in this work we propose and motivate a method of indirectly measuring this convergence when applying the GC algorithm to some representative Hamiltonian systems of two and three degrees of freedom, such that the transient phases can be terminated once all the relevant subspaces are sufficiently close to be considered converged. Furthermore, we propose a minor adaptation of the GC algorithm that improves the accuracy of the so-called center subspaces when computed over long time intervals. Our numerical results for this investigation are presented in Sect.~\ref{sec:results}. Finally, our conclusions are given in Sect.~\ref{sec:conclusions} and our recommendations for users of the GC algorithm are summarized.


\section{Theoretical aspects and numerical methods}
\label{sec:theory}

In this paper, we focus our attention on continuous time dynamical systems with time invertible dynamics in order to guarantee the existence of the so-called Oseledets splitting (see Sect.~\ref{sec:clv}). Considering such a dynamical system with an $N$-dimensional ($N$-D) phase space, we denote by $\bs x$ a point in this space. A so-called deviation vector $\bs w$ is any element of the $N$-D tangent space of the system at $\bs x$. By interpreting these deviation vectors as arbitrarily small perturbations to $\bs x$, we track their time evolution  in order to study the stability properties of the dynamical system. A deviation vector $\bs w$ is dynamically evolved forward by $t$ units of time via multiplication with the $N\times N$ fundamental matrix $M(\bs x,t)$ of the system (see e.g.~\cite{Skokos2010}).

\subsection{Covariant Lyapunov vectors}
\label{sec:clv}

Following the presentations of \cite{KuptsovParlitz2012,GinelliEtAl2013}, we provide here a brief overview of the theory of CLVs and their computation via the GC algorithm. To assist the reader with the technical definitions that follow, we provide a simple example of a dynamical system in Appendix~\ref{sec:appendix}, offering a more concrete illustration of these quantities.

It follows from the multiplicative ergodic theorem \cite{Oseledets1968} that, for almost any initial condition $\bs x$, the limit
\begin{align}
    \lambda^{\pm}(\bs w) := \lim_{t\to\pm\infty}\frac{1}{t}\ln\frac{\|M(\bs x,t)\bs w\|}{\|\bs w\|}\label{eq:growth}
\end{align}
exists for all non-zero deviation vectors $\bs w$, and furthermore that $\lambda^{\pm}(\bs w)$ takes one of $m\le N$ distinct values,
\begin{align}
    \lambda_1>\lambda_2>\cdots>\lambda_m,\label{eq:spec}
\end{align}
known as LEs. In particular, both forward time and backward time filtrations of the tangent space $T_{\bs x}\mathcal{M}$ exist, given respectively by
\begin{align}
    \begin{split}
        T_{\bs x}\mathcal M&=\Gamma_1^+(\bs x)\supset\Gamma_2^+(\bs x)\supset\cdots\supset\Gamma_m^+(\bs x)\supset\Gamma_{m+1}^+(\bs x):=\{0\},\\
        T_{\bs x}\mathcal M&=\Gamma_m^-(\bs x)\supset\Gamma_{m-1}^-(\bs x)\supset\cdots\supset\Gamma_1^-(\bs x)\supset\Gamma_0^-(\bs x):=\{0\},
    \end{split}\label{eq:filter4}
\end{align}
such that $\lambda^{\pm}(\bs w)=\lambda_i$ if $\bs w\in\Gamma_i^{\pm}(\bs x)\setminus\Gamma_{i\pm1}^{\pm}(\bs x)$, where $i=1,\dots,m$. Furthermore, a splitting of the tangent space can be constructed:
\begin{align}
    T_{\bs x}\mathcal M=\Omega_1(\bs x)\oplus\Omega_2(\bs x)\oplus\cdots\oplus\Omega_m(\bs x), \label{eq:splitting}
\end{align}
where $\Omega_i(\bs x) := \Gamma_i^+(\bs x) \cap \Gamma_i^-(\bs x)$. The filtration subspaces can be decomposed in terms of the splitting subspaces as follows:
\begin{align}
	\begin{split}
		\Gamma_i^-(\bs x) &= \Omega_1(\bs x)\oplus\Omega_2(\bs x)\oplus\cdots\oplus\Omega_i(\bs x),\\
		\Gamma_i^+(\bs x) &= \Omega_i(\bs x)\oplus\Omega_{i+1}(\bs x)\oplus\cdots\oplus\Omega_m(\bs x),
	\end{split}\label{eq:relation}
\end{align}
\cite{Noethen2019a}. If $\Omega_i$ corresponds to a zero LE $\lambda_i=0$, then we call $\Omega_i$ the center subspace. A set of $N$ linearly independent unit vectors that span the splitting \eqref{eq:splitting} are known as CLVs. Note that if $\dim\Omega_i=1$, then $\Omega_i$ contains only one CLV, which is unique up to a sign, otherwise any $\dim\Omega_i$ linearly independent unit vectors in $\Omega_i$ can be chosen arbitrarily as the CLVs of $\Omega_i$.

\begin{figure}[htbp]
    \centering
    \includegraphics[width=\linewidth]{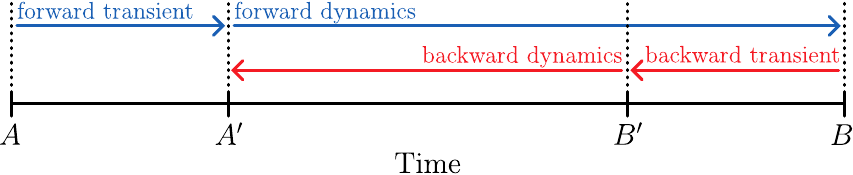}
    \caption{Diagram depicting four phases of the GC algorithm.}
    \label{fig:phases}
\end{figure}

Generically, a randomly chosen subspace $G_i$ of $T_{\bs x}\mathcal{M}$ satisfying $\dim G_i=\dim\Gamma_i^-$ will converge exponentially fast to the filtration subspace $\Gamma_i^-$ under the forward evolution of the tangent dynamics, i.e.\ $G_i\to\Gamma_i^-$ as $t\to\infty$ for each $i=1,\dots,m$. As shown in \cite{Noethen2019,Noethen2019a}, these exponential rates of convergence are given by the spectral gaps $\lambda_i-\lambda_{i+1}$. Furthermore, a generic subspace $C_i$ of $\Gamma_i^-$ satisfying $\dim C_i=\dim \Omega_i$ will converge exponentially fast to the splitting subspace $\Omega_i$ under backward evolution, i.e.\ $C_i\to\Omega_i$ as $t\to-\infty$ for each $i=1,\dots,m$. At least in the case when each $\Omega_i$ is 1-D, these exponential convergence rates are $\lambda_{i-1}-\lambda_i$ \cite{GinelliEtAl2013}. The GC algorithm computes the first $k\leq N$ CLVs by taking advantage of this behavior of the tangent dynamics. The four phases of the algorithm are depicted in Fig.~\ref{fig:phases} and can be summarized as follows:
\begin{enumerate}
    \item \textbf{Forward transient phase.} Initialize a set of $k\leq N$ random deviation vectors. Evolve these vectors forward in time until the subspaces $G_i$ they span converge to $\Gamma_i^-$.
    \item \textbf{Forward dynamics phase.} Further evolve these deviation vectors forward in time until the filtration subspaces $\Gamma_i^-$ have been accurately estimated over a desired time interval.
    \item \textbf{Backward transient phase.} Initialize a set of $k\leq N$ random deviation vectors which span the estimated filtration subspaces $\Gamma_i^-$. Evolve these vectors backward in time until the subspaces $C_i$ they span converge to $\Omega_i$.
    \item \textbf{Backward dynamics phase.} Further evolve these deviation vectors backward in time until the splitting subspaces $\Omega_i$ have been accurately estimated over a desired time interval.
\end{enumerate}
Note, however, that our summary of the GC algorithm omits important implementation details that ensure the numerical stability of the GC algorithm---such as the repeated orthonormalization of deviation vectors during the forward evolution parts of the algorithm---because such details are covered extensively elsewhere, see e.g.\ \cite{GinelliEtAl2013,PikovskyPoliti2016,duPlessis2024}.

The purpose of the two transient phases of the GC algorithm is to give the computed subspaces sufficient time to converge to the relevant filtration/splitting subspaces: $G_i\to\Gamma_i^-$ during the forward transient phase, and $C_i\to\Omega_i$ during the backward transient phase. The CLVs are computed between times $A'$ and $B'$, indicated in Fig.~\ref{fig:phases}, during which we can rely on the accuracy of these CLVs. However, if the length of the transient phases is too short, then the computed subspaces will not have converged yet, so the accuracy of the computed CLVs will be poor for a short time after $A'$ and before $B'$ while these subspaces are still converging (which they continue doing as they dynamically evolve). Avoiding this issue is the motivation behind the methods we introduce later on in Sect.~\ref{sec:converge}, which ensure that the transient phases for a particular computation are only terminated once the relevant subspaces have converged.

Since the number $m$ of distinct LEs is not known a priori, numerical methods for evaluating LEs \cite{BenettinEtAl1980a,BenettinEtAl1980,Skokos2010} typically compute a spectrum of $N$ possibly non-distinct (or degenerate) LEs,
\begin{align}
    \chi_1\ge\chi_2\ge\cdots\ge\chi_N,\label{eq:nonspec}
\end{align}
where the value of each distinct LE $\lambda_i$ from \eqref{eq:spec} appears precisely $\dim\Omega_i$ times in the non-distinct spectrum \eqref{eq:nonspec}. Since the multiplicity of each LE in \eqref{eq:nonspec} is $\dim\Omega_i$, the number and dimensions of the splitting subspaces $\Omega_i$ is straightforward to determine from knowledge of the degeneracies in the LE spectrum and their multiplicities. Additionally, the number and dimensions of the filtration subspaces $\Gamma_i^-$ are easily determined using \eqref{eq:relation}. Since our numerical investigation focuses on autonomous Hamiltonian systems, we note that the non-distinct LE spectrum for such systems has the following symmetry: $\chi_j=-\chi_{N-j+1}$ for $j=1,\dots,N$. Furthermore, for any differentiable flow on a compact manifold, an orbit which does not terminate at a fixed point has at least one vanishing LE \cite{BenettinEtAl1980a,Haken1983}. Therefore, the middle pair of LEs for an autonomous Hamiltonian system is zero, i.e.\ $\chi_n=\chi_{n+1}=0$, where $n=N/2$ is the number of degrees of freedom ($N$ is even for such systems).

\subsection{Measuring the convergence of subspaces}
\label{sec:converge}

Following \cite{GolubVanLoan1996,YeLim2016}, we define the distance $\Delta\in[0,1]$ (also called the aperture \cite{AkhiezerGlazman1993} or gap \cite{DezaDeza2009}) between equidimensional subspaces $A$ and $B$ as
\begin{align}
	\Delta[A,B]:=\sin\theta_{\mathrm{max}},\label{eq:distance}
\end{align}
where $\theta_{\mathrm{max}}\in[0,\pi/2]$ is the largest principal angle between $A$ and $B$. In Appendix~\ref{sec:appendix2}, we include a brief discussion of principal angles, including their definition, computation, and two examples of the principal angles between low-dimensional subspaces. We will use this measure of distance in order to quantify the level of convergence between various subspaces computed via the GC algorithm. For convenience, we also define the distance $\Delta[\bs a,\bs b]$ between vectors $\bs a$ and $\bs b$ as $\Delta[\spn(\bs a),\spn(\bs b)]$.

In our study, we measure the convergence of estimates $G_i$ of the filtration subspaces $\Gamma_i^-$ during the forward transient phase of the GC algorithm. In order to specify the time interval over which a subspace $G_i$ is computed, we use the notation $G_i(t',t'')$ to denote an estimate of $\Gamma_i^-$ at time $t''$ using deviation vectors which are chosen at random at time $t'$ and evolved forward to time $t''$ ($t''>t'$). We write $G_i$ and $\tilde G_i$ to distinguish between two estimates of the same subspace which use different sets of random initial deviation vectors. At the end of the forward dynamics phase, the backward transient phase begins by evolving backward from that point in time; we parametrize this point in time by $t_b=0$, where $t_b$ is a \textit{backward} time variable that increases during backward evolution, thus avoiding the inconvenience of working with a decreasing time variable (such as $t$) during the backward transient phase. Similarly to the forward transient phase, we denote an estimate of the splitting subspace $\Omega_i$ at backward time $t_b''$ by $C_i(t_b',t_b'')$, which is computed using deviation vectors initialized at backward time $t_b'$ and evolved backward in time to $t_b''$ ($t_b''>t_b'$). $C_i$ and $\tilde C_i$ are two estimates of the same subspace using different random sets of initial deviation vectors.

In Sects.~\ref{sec:m2} and \ref{sec:m3}, we introduce two different computational methods of comparing the convergence rates of subspaces via the evaluation of their distance $\Delta$. The direct method measures the convergence between a pre-computed accurate estimate of a desired subspace and an estimate computed during one of the transient phases of the GC algorithm, thus directly testing the accuracy of the subspaces computed during the transient phases. The indirect method, on the other hand, measures the convergence between two independent estimates of the same subspace, providing indirect evidence of their convergence to the desired subspace by their convergence to each other. It is worth mentioning that an entirely different numerical method of directly measuring convergence during the transient phases of the GC algorithm was demonstrated in \cite{Noethen2019a}, but we exclude this method from our discussion since it only measures convergence of the relevant subspaces at a single point in phase space and involves many more numerical integrations than the methods we discuss here, so we rather opt for more efficient alternatives which can be at least in part computed during the transient phases of the GC algorithm.

\subsubsection{The direct method}
\label{sec:m2}

\begin{figure}[htbp]
    \centering
    \includegraphics[width=\linewidth]{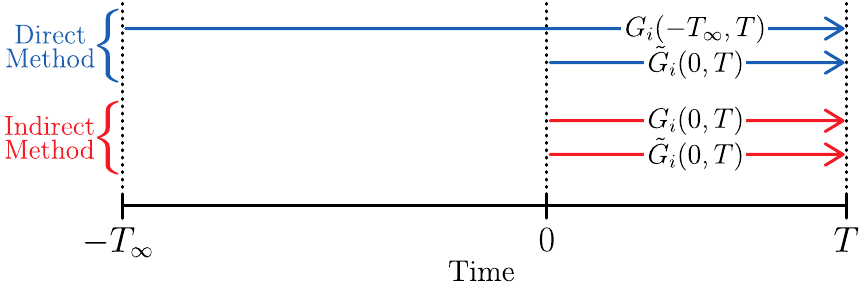}
    \caption{Diagram depicting the numerical integrations required to implement the direct and indirect methods discussed in Sects.~\ref{sec:m2} and \ref{sec:m3} (respectively) for measuring the convergence between estimates $G_i$ and $\tilde G_i$ of the filtration subspaces $\Gamma_i^-$ during the forward transient phase of the GC algorithm, but both methods can also be applied to the backward transient phase. The time $T$ denotes a time in the future, while $-T_{\infty}$ denotes a distant time in the past.}
    \label{fig:diagram}
\end{figure}

The first method we discuss begins by pre-computing the filtration subspaces $\Gamma_i^-$ ($i=1,\dots,m)$ at time $t=0$ as accurately as possible by evolving a set of deviation vectors over a very long time interval of length $T_{\infty}$. For sufficiently long $T_{\infty}$, we can assume that $G_i(-T_{\infty},0)$ is an accurate estimate of $\Gamma_i^-$ at time $t=0$ and, by continuing to evolve this subspace up to some final time $T>0$, the subspace $G_i(-T_{\infty},t)$ accurately estimates $\Gamma_i^-$ over the entire interval $t\in[0,T]$. After this pre-computation, we begin the forward transient phase of the GC algorithm at $t=0$ by computing a new estimate $\tilde G_i$ of $\Gamma_i^-$ over the time interval $t\in[0,T]$. During the forward transient phase, we also compute the distance $\Delta[G_i(-T_{\infty},t),\tilde G_i(0,t)]$, which directly checks the accuracy of the subspaces $\tilde G_i$ by comparing them with their corresponding accurate estimates $G_i$ of $\Gamma_i^-$. This method is represented by the blue arrows in Fig.~\ref{fig:diagram}. This method is similarly applied during the backward transient phase by computing the backward time $t_b$ evolution of $\Delta[C_i(-T_{\infty},t_b),\tilde C_i(0,t_b)]$, where the subspaces $\tilde C_i(0,t_b)$ computed during the transient phase are compared with the accurate estimates $C_i(-T_{\infty},t_b)$ of $\Omega_i$ at backward time $t_b$.

This method directly checks the accuracy of the various subspaces computed during the transient phases of the GC algorithm. However, it suffers from one drawback: the length $T_{\infty}$ must be arbitrarily decided a priori such that it is long enough to guarantee convergence of the relevant subspaces yet not too long that it becomes infeasible to compute the accurate subspace estimates over this interval. We therefore introduce a more efficient and pragmatic approach in Sect.~\ref{sec:m3}.

\subsubsection{The indirect method}
\label{sec:m3}

The second method we discuss was introduced in \cite{duPlessis2024}, inspired by the Smaller Alignment Index (SALI) chaos indicator \cite{Skokos2001,SkokosEtAl2003,SkokosEtAl2004,SkokosManos2016}, which detects chaos by measuring the alignment of two randomly chosen deviation vectors as they dynamically evolve. Unlike the previous method that directly measures the level of convergence of computed estimates of the relevant subspaces by comparing them to considered accurate estimates of those subspaces obtained from the evolution of deviation vectors for a long transient phase of length $T_{\infty}$, this method indirectly measures convergence by comparing two independent estimates of the same subspaces over the same interval. To do this, we compute two estimates $G_i(0,t)$ and $\tilde G_i(0,t)$ of the filtration subspace $\Gamma_i^-$ on the interval $t\in[0,T]$ and find their distance $\Delta[G_i(0,t),\tilde G_i(0,t)]$. The two numerical integrations required for this method are represented by the red arrows in Fig.~\ref{fig:diagram}. Not only is this method faster than the direct method, but it has no need for the arbitrary choice of a sufficiently long $T_{\infty}$. Note though that since the forward transient phase of the GC algorithm only requires the computation of the estimates $G_i$, the additional computations of $\tilde G_i$ mean that applying this method will effectively double the CPU time required for the transient phases of the algorithm.

But is this indirect method reliable? In particular for the forward transient phase, is it true that if $G_i$ and $\tilde G_i$ have converged up to some desired threshold that they have both similarly converged to $\Gamma_i^-$? Since $\Delta$ is a metric \cite{Noethen2019,Noethen2019a}, it obeys the triangle inequality $\Delta[G_i,\tilde G_i]\leq\Delta[G_i,\Gamma_i^-]+\Delta[\tilde G_i,\Gamma_i^-]$, from which we see that if $G_i$ and $\tilde G_i$ are each close to $\Gamma_i^-$, then $G_i$ and $\tilde G_i$ must be similarly close to each other. However, the converse is not always true, since particular deviation vectors can be contrived such that $G_i$ and $\tilde G_i$ converge to each other at faster rates than they converge to $\Gamma_i^-$ (see e.g.\ Sect.~2.6 of \cite{duPlessis2024}). We nevertheless expect that such special choices of deviation vectors are extremely unlikely (forming a probability zero subset of all possible choices) and that this method is generically reliable for almost all choices of deviation vectors. We provide numerical evidence for the reliability of this method in Sect.~\ref{sec:results} by comparing it with the direct method and showing that the results are extremely similar.

\subsection{The considered Hamiltonian systems}
\label{sec:Hams}

We present here the two autonomous Hamiltonian systems that we use to demonstrate and compare the direct and indirect methods presented in Sect.~\ref{sec:converge}. Choosing an example chaotic orbit for each, we discuss the degeneracies in the LE spectrum and thus the dimensionality of the filtration and splitting subspaces to be computed when we apply the GC algorithm.

\subsubsection{The H\'enon-Heiles system}
\label{sec:hh}

The first system we use in our investigation is the well-known H\'enon-Heiles system \cite{HenonHeiles1964}, which has the Hamiltonian
\begin{align}
    H_2=\frac12(p_x^2+p_y^2)+\frac12(x^2+y^2)+x^2y-\frac13y^3,\label{eq:hamiltonian}
\end{align}
with $x$, $y$ being generalized coordinates and $p_x$, $p_y$ respectively the conjugate momenta. This system has two degrees of freedom and a phase space of dimension $N=4$. We consider a chaotic orbit in our numerical investigation, so the maximum LE $\chi_1$ (or $\lambda_1$) is positive. The H\'enon-Heiles system is an autonomous Hamiltonian system with a 4-D phase space, so its $N=4$ non-distinct LEs \eqref{eq:nonspec} for a chaotic orbit are given by $\chi_1>0$, $\chi_2=\chi_3=0$, and $\chi_4=-\chi_1<0$. By combining the repeated LEs, we infer that it has $m=3$ distinct LEs \eqref{eq:spec} given by $\lambda_1>0$, $\lambda_2=0$, and $\lambda_3=-\lambda_1<0$.

Since $\lambda_2$ has a multiplicity of 2 (it appears twice in the non-distinct LE spectrum), the splitting subspace $\Omega_2$ corresponding to $\lambda_2$ has dimension 2, while the remaining splitting subspaces $\Omega_1$ and $\Omega_3$ are each 1-D since they correspond to non-degenerate LEs. Therefore, it follows from \eqref{eq:relation} that the dimensions of the filtration subspaces are $\dim\Gamma_1^-=1$, $\dim\Gamma_2^-=1+2=3$, and $\dim\Gamma_3^-=1+2+1=4$. The dimensions of the filtration/splitting subspaces inform us of which subspaces computed from the deviation vectors during the transient phases of the GC algorithm will converge to the relevant filtration/splitting subspaces. In particular, if we denote the $N=4$ deviation vectors in the forward transient phase by $\hat{\bs g}_1$, $\hat{\bs g}_2$, $\hat{\bs g}_3$, and $\hat{\bs g}_4$, then we let $G_1=\spn(\hat{\bs g}_1)$, $G_2=\spn(\hat{\bs g}_1,\hat{\bs g}_2,\hat{\bs g}_3)$, and $G_3=\spn(\hat{\bs g}_1,\hat{\bs g}_2,\hat{\bs g}_3,\hat{\bs g}_4)$ so that $\dim G_i=\dim\Gamma_i^-$ for each $i=1,2,3$, and hence $G_i\to\Gamma_i^-$ as $t\to\infty$. For the backward transient phase, each initial deviation vector $\hat{\bs c}_j$ is randomly sampled from the subspace $\spn(\{\hat{\bs g}_k\}_{k\leq j})$ for $j=1,2,3,4$; we then define the subspaces $C_1=\spn(\hat{\bs c}_1)$, $C_2=\spn(\hat{\bs c}_2,\hat{\bs c}_3)$, and $C_3=\spn(\hat{\bs c}_4)$ so that $C_i\subseteq\Gamma_i^-$ and $\dim C_i=\dim\Omega_i$ for each $i=1,2,3$, and hence $ C_i\to\Omega_i$ as $t_b\to\infty$.

\subsubsection{A Hamiltonian system with three degrees of freedom}
\label{sec:hh3D}

The second system we consider in our numerical investigation is an autonomous Hamiltonian system with three degrees introduced in \cite{ContopoulosEtAl1978} that exhibits both regular and chaotic dynamics. The Hamiltonian for this system is given by
\begin{align}
    H_3 = \sum_{i=1}^3\frac{\omega_i}{2}(p_i^2 + q_i^2) + q_1^2(q_2+q_3)\label{eq:3d_hamiltonian}
\end{align}
where $\omega_i=\sqrt{i}$ for $i=1,2,3$, and $q_i$ and $p_i$ are the respective position and momentum coordinates. The dimension of this system's phase space is $N=6$. We consider a chaotic orbit with distinct LEs except for the unavoidable middle pair of zero LEs, i.e.\ $\chi_1>\chi_2>\chi_3=0=\chi_4>\chi_5>\chi_6$ (with $\chi_5=-\chi_2$ and $\chi_6=-\chi_1$), so it has five distinct LEs $\lambda_1>\lambda_2>\lambda_3>\lambda_4>\lambda_5$ where $\lambda_3=0$ has multiplicity $2$ and the rest have a multiplicity of $1$.

Due to the structure of the LE spectrum, the splitting subspaces are all 1-D, except for $\Omega_3$ which has dimension 2. It follows from \eqref{eq:relation} that the filtration subspaces have dimensions $\dim\Gamma_1^-=1$, $\dim\Gamma_2^-=2$, $\dim\Gamma_3^-=4$, $\dim\Gamma_4^-=5$, and $\dim\Gamma_5^-=6$. If we denote the $N=6$ deviation vectors in the forward transient phase by $\hat{\bs g}_j$ for $j=1,\dots,6$, then we let $G_1=\spn(\hat{\bs g}_1)$, $G_2=\spn(\hat{\bs g}_1,\hat{\bs g}_2)$, $G_3=\spn(\hat{\bs g}_1,\hat{\bs g}_2,\hat{\bs g}_3,\hat{\bs g}_4)$, etc.\ so that $\dim G_i=\dim\Gamma_i^-$ and hence $G_i\to\Gamma_i^-$ as $t\to\infty$ for $i=1,\dots,5$. For the backward transient phase, each initial deviation vector $\hat{\bs c}_j\in\spn(\{\hat{\bs g}_k\}_{k\leq j})$ for $j=1,\dots,6$, and we define the subspaces $C_1=\spn(\hat{\bs c}_1)$, $C_2=\spn(\hat{\bs c}_2)$, $C_3=\spn(\hat{\bs c}_3,\hat{\bs c}_4)$, $C_4=\spn(\hat{\bs c}_5)$, and $C_5=\spn(\hat{\bs c}_6)$ so that $C_i\subseteq\Gamma_i^-$ and $\dim C_i=\dim\Omega_i$, and hence $ C_i\to\Omega_i$ under backward evolution.


\section{Numerical results}
\label{sec:results}

In our investigations, we examine several chaotic orbits in both systems, but show results for only one representative orbit for each Hamiltonian model. We first present our results for the H\'enon-Heiles system \eqref{eq:hamiltonian} and then for the system with three degrees of freedom \eqref{eq:3d_hamiltonian}. For our numerical integrations, we implement the ABA864 symplectic integrator of order 4 \cite{BlanesEtAl2013}, which has been shown to be very efficient \cite{DanieliEtAl2019}. In particular, we use a time step $\tau=0.025$ for the H\'enon-Heiles system \eqref{eq:hamiltonian} and $\tau=0.02$ for system \eqref{eq:3d_hamiltonian}, which we find ensures that the relative energy error is always less than $10^{-10}$.

\subsection{Results for the H\'enon-Heiles system}
\label{sec:hh_results}

For our computations, we consider the H\'enon-Heiles system with energy $H_2=1/6$ and initial condition
\begin{align}
    (x,y,p_y)=(0,0.54,-0.06),\label{eq:ic}
\end{align}
where $p_x\geq0$ is calculated using \eqref{eq:hamiltonian}. In order to check that this orbit is indeed chaotic, we estimate the mLE using the so-called `standard method' \cite{BenettinEtAl1980,BenettinEtAl1980a} (which we implement using the pseudocode presented in \cite{Skokos2010}) by computing the finite-time mLE over an interval $t\in[0,10^7]$, and we find that $\chi_1 \approx 0.127$. Therefore, the orbit is chaotic and the discussion in Sect.~\ref{sec:hh} around the LE spectrum and the dimensions of the filtration/splitting subspaces applies here.

\subsubsection{Forward transient phase}
\label{sec:forward}

\begin{figure}[htbp]
    \centering
    \includegraphics[width=0.8\linewidth]{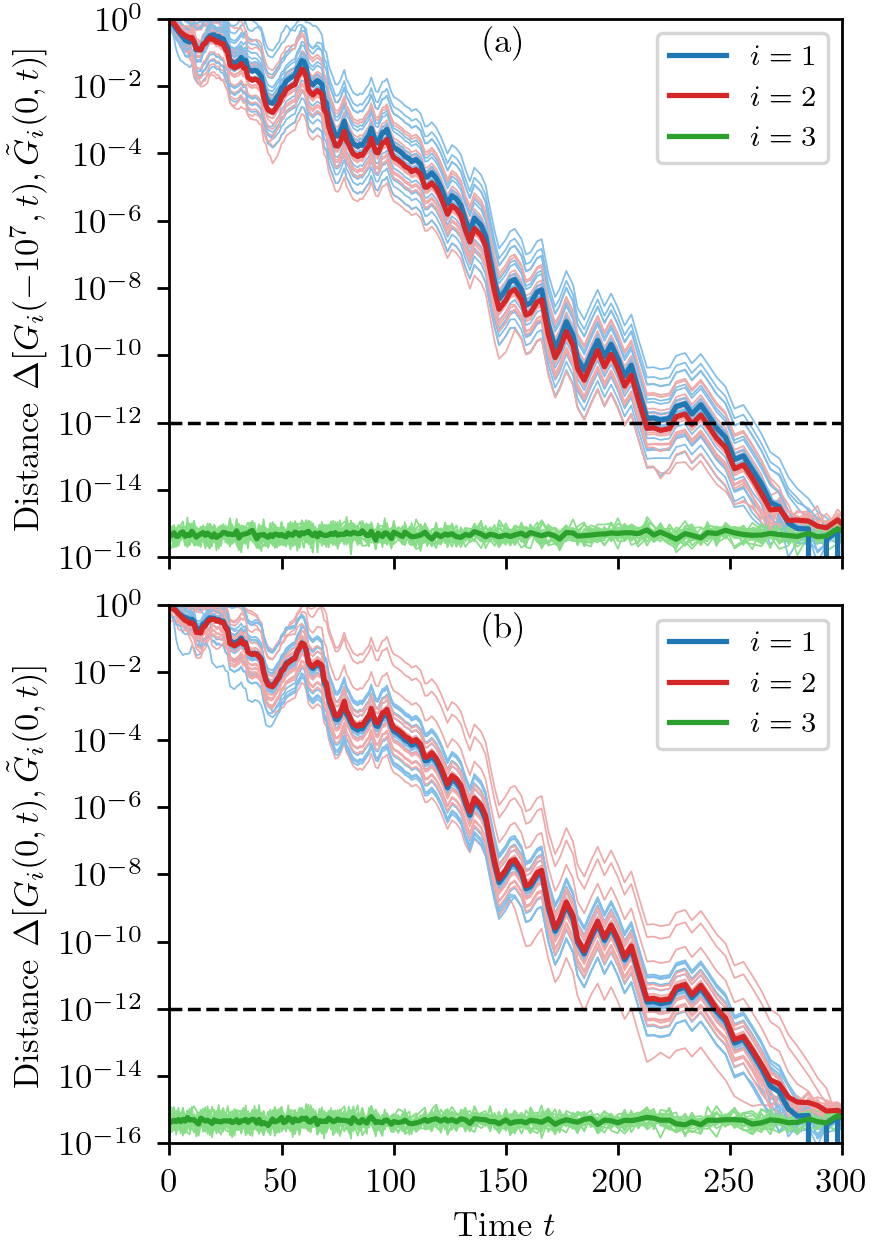}
    \caption{The time evolution of the distance $\Delta$ \eqref{eq:distance} between estimates of $\Gamma_i^-$ for $i=1,2,3$, computed using the (a) direct and (b) indirect methods of Sect.~\ref{sec:converge} during the forward transient phase of the GC algorithm for a chaotic orbit of the Hénon-Heiles system \eqref{eq:hamiltonian} with initial condition \eqref{eq:ic} and $H_2=1/6$. Each thin curve represents results obtained for one of 20 sets of random initial deviation vectors, while the thicker curves represent the average $\Delta$ values in log scale over these 20 simulations. Note that the thick blue ($i=1$) and red ($i=2$) curves practically overlap each other. Both panels are in log-linear scale, and the black dashed lines represent $\Delta=10^{-12}$.}
    \label{fig:forward}
\end{figure}

We now apply the forward transient phase of the GC algorithm to initial condition \eqref{eq:ic} corresponding to a chaotic orbit of the H\'enon-Heiles system and compute the time evolution of estimates of the filtration subspaces $\Gamma_i^-$ for $i=1,2,3$. For the interval lengths $T_{\infty}$ and $T$ described in Fig.~\ref{fig:diagram}, we use $T_{\infty}=10^7$ and $T=300$ time units. The time evolution of the distance $\Delta$ between independent estimates $G_i$ and $\tilde G_i$ of $\Gamma_i^-$ computed using the direct and indirect methods are shown in panels (a) and (b) of Fig.~\ref{fig:forward}, respectively, where we present results for 20 simulations that each use a different set of random initial deviation vectors. The black dashed lines in Fig.~\ref{fig:forward} represent an arbitrary small threshold of $\Delta=10^{-12}$; when two subspaces come close enough to each other that the distance $\Delta$ between them crosses below this threshold, we deem the two subspaces to have converged with each other. We see from Fig.~\ref{fig:forward} that for both methods, the distance $\Delta$ between $G_i$ and $\tilde G_i$ decreases at an exponential rate for $i=1,2$, and the first time at which $\Delta$ for all relevant subspaces crosses the $10^{-12}$ threshold lies in the interval $t\in[200,270]$ for each simulation. We note that for $i=3$, no convergence phase is expected because, as discussed in Sect.~\ref{sec:hh}, $\dim\Gamma_3^-=4$ and hence $\dim G_3=\dim \tilde G_3=4$, so these subspaces of the 4-D tangent space must all coincide with the entire tangent space; since no convergence is required, this explains why $\Delta[G_3,\tilde G_3]$ remains approximately constant at a value smaller than $10^{-15}$, which is near the computational limit of double precision used in our numerics.

We observe from Fig.~\ref{fig:forward} that for each simulation, all relevant pairs of subspaces converge by some time $t\in[200,270]$ once the threshold is crossed. It is important to note, however, that while the long-time average convergence rates are related to the gaps in the LE spectrum, these rates can fluctuate in the short term and result in significantly different convergence times for different initial conditions. For example, we have repeated these simulations for a so-called sticky orbit where all the relevant subspaces took almost $1000$ time units to converge. We therefore recommend that the forward transient phase of the GC algorithm be stopped not at some arbitrarily chosen time $T$ but instead as soon as the distance $\Delta$ between computed subspaces $G_i$ and $\tilde G_i$ has decreased below the threshold for every $i=1,2,3$ for that particular simulation; this ensures the accuracy of the computed filtration subspaces for that simulation while saving CPU time by not using an unnecessarily long transient phase. We also note that there is no significant difference between our results for the direct and indirect methods, so we recommend the simpler and faster indirect method which can be computed entirely during the forward transient phase of the GC algorithm.

\subsubsection{Backward transient phase}
\label{sec:back}

After the forward transient phase, we evolve the system further forward in time by approximately $10^7$ time units for the forward dynamics phase of the GC algorithm. From this later point in time, we begin the backward transient phase and apply the direct and indirect methods to measure the convergence of the splitting subspace estimates. For the direct method, we again use $T_{\infty}=10^7$ as the interval length over which the splitting subspaces are pre-computed. For reasons which will become clear, we perform the backward transient phase over an unnecessarily long backward time interval of length $T=10^7$ for 20 simulations that each use different sets of random initial deviation vectors.

\begin{figure}[p]
    \centering
    \includegraphics[width=0.8\linewidth]{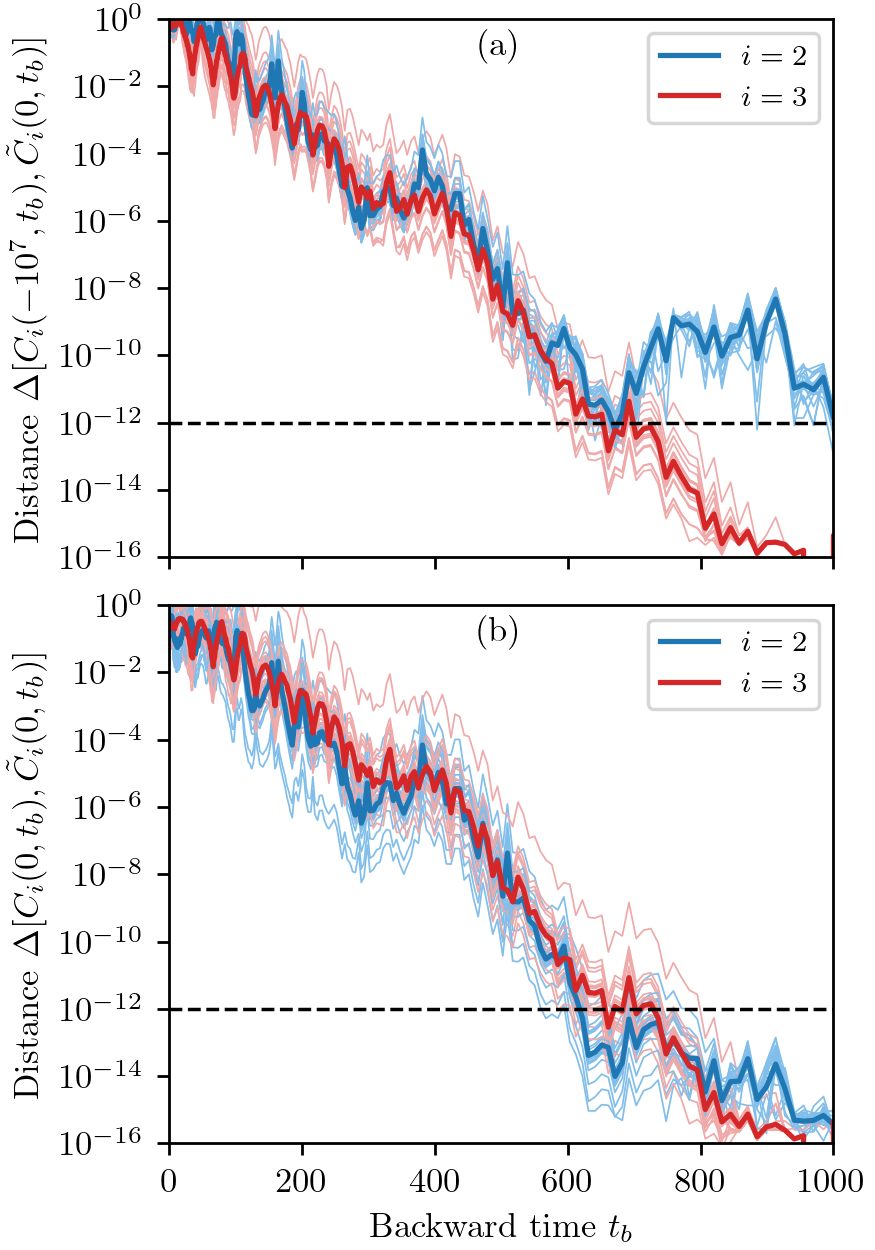}
    \caption{The backward time evolution of the distance $\Delta$ between estimates of $\Omega_i$ for $i=2,3$, computed using the (a) direct and (b) indirect methods of Sect.~\ref{sec:converge} during the first $1000$ time units of the backward transient phase of the GC algorithm for the same orbit used in Fig.~\ref{fig:forward}. Each thin curve represents results obtained for one of 20 sets of random initial deviation vectors, while the thicker curves represent the average $\Delta$ values in log scale over these 20 simulations. Both panels are in log-linear scale, and the black dashed lines represent $\Delta=10^{-12}$.}
    \label{fig:back_lin}
\end{figure}

\begin{figure}[p]
    \centering
    \includegraphics[width=0.8\linewidth]{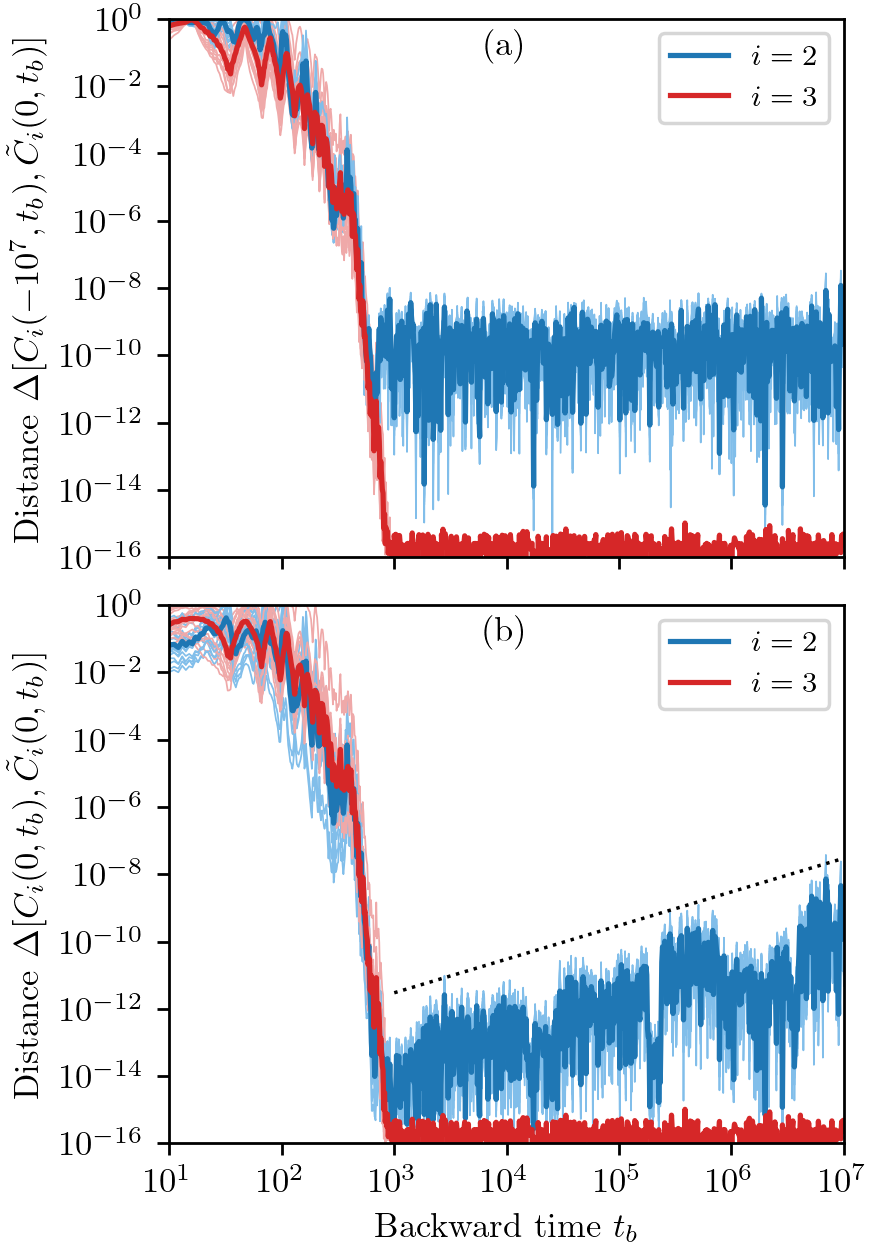}
    \caption{Similar to Fig.~\ref{fig:back_lin}, where panels (a) and (b) correspond between the figures. This figure, however, is in log-log scale and the backward time evolution of $\Delta$ over the entire backward transient interval of $10^7$ is shown here. The black dotted line in (b) denotes a function $\propto t_b$.}
    \label{fig:back_log}
\end{figure}

Figure~\ref{fig:back_lin} shows the backward time evolution of the distance $\Delta$ between estimates $C_i$ and $\tilde C_i$ of $\Omega_i$ for $i=2,3$ computed over the first $1000$ time units during the backward transient phase using the direct method in panel (a) and the indirect method in panel (b). Note that the $i=1$ case is omitted from the figure since $C_1$ and $\tilde C_1$ are both 1-D subspaces of the 1-D subspace $G_1$ (see Sect.~\ref{sec:hh}), so these subspaces all coincide and so the distance $\Delta$ between any pair of them is zero. Now, we see from both panels of Fig.~\ref{fig:back_lin} that at first $\Delta$ decreases at an exponential rate for both $i=2$ and $i=3$; however, in Fig.~\ref{fig:back_lin}(a) for the $i=2$ case, we see that the center subspace estimates $C_2$ and $\tilde C_2$ fail to reliably converge up to the desired threshold of $\Delta=10^{-12}$ for some simulations during the time interval shown. It becomes clear that this issue affects both methods when we zoom out and analyze the entire $10^7$ backward transient phase, the results for which are plotted in log-log scale in Fig.~\ref{fig:back_log}. We see from Fig.~\ref{fig:back_log}(a) for the $i=2$ case of the direct method that $\Delta$ only decreases to about $10^{-10}$ and fluctuates around this value for the remainder of the backward integration, while from Fig.~\ref{fig:back_log}(b) for the indirect method we see that $\Delta$ initially decreases to about $10^{-14}$ before increasing linearly with backward time $t_b$. This increasing disagreement between two independent estimates of the same subspace using the indirect method suggests that the application of the GC algorithm to compute this 2-D center subspace in the H\'enon-Heiles system is numerically unstable over long backward time intervals. It is important to note that this issue is not merely an artifact of our very long backward transient phase, since this inaccuracy would also appear and persist during the backward dynamics phase.

\begin{figure}[htbp]
    \centering
    \includegraphics[width=0.8\linewidth]{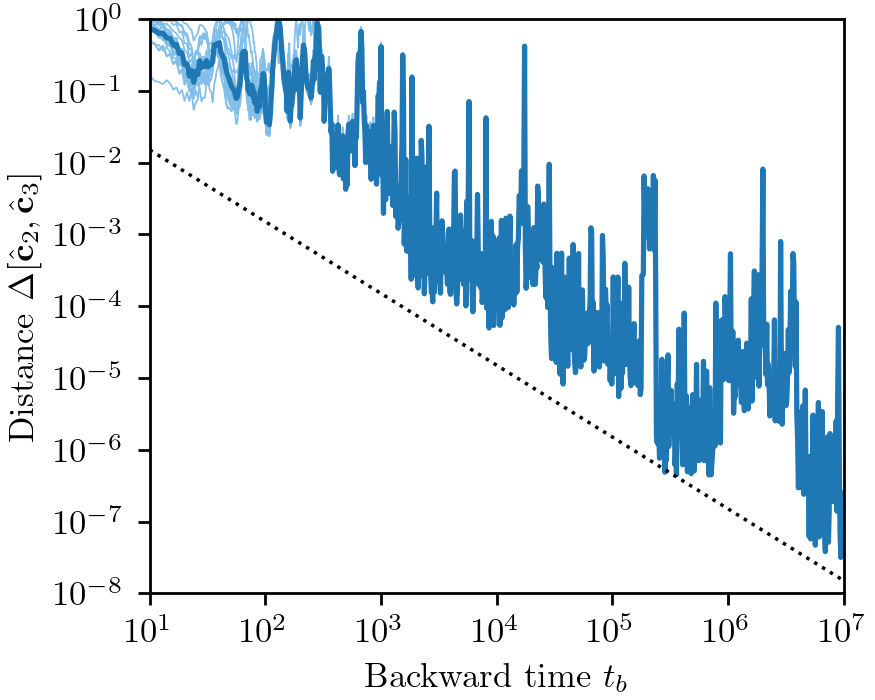}
    \caption{The backward time evolution of the distance $\Delta$ between the two linearly independent CLV estimates $\hat{\bs c}_2$ and $\hat{\bs c}_3$ in the computed center subspace $C_2$. The black dotted line denotes a function $\propto t_b^{-1}$. The figure is in log-log scale.}
    \label{fig:center}
\end{figure}

The poor convergence between estimates of the center subspace observed in Fig.~\ref{fig:back_log} for both methods can be explained by the dynamics of the center subspace itself. Figure~\ref{fig:center} shows the backward time evolution of the distance $\Delta$ between linearly independent CLV estimates $\hat{\bs c}_2$ and $\hat{\bs c}_3$ that span the subspace $C_2$, which converges to the center subspace $\Omega_2$. We see from this figure that over long backward times, the distance between these vectors evolves according to $\Delta[\hat{\bs c}_2,\hat{\bs c}_3]\propto t_b^{-1}$. This observation suggests that any two estimates of CLVs in the center subspace computed via the GC algorithm slowly align/anti-align during backward evolution, an idea which is studied in more detail in \cite{duPlessis2024}.

In light of this convergence between center subspace CLVs, we offer the following explanations for the poor convergence of center subspace estimates seen in Fig.~\ref{fig:back_log} for each of the two methods:
\begin{description}
    \item[Direct method.] The long-time estimate $C_2(-10^7,t_b)$ of the center subspace becomes imprecise over the long $10^7$ interval, during which the two CLV estimates which span $C_2$ become significantly aligned/anti-aligned, resulting in poor numerical accuracy when computing $C_2$. Therefore, after an initial exponentially fast decay of the distance $\Delta$ between $C_2(-10^7,t_b)$ and $\tilde C_2(0,t_b)$, this distance then becomes dominated by the inaccuracy of $C_2(-10^7,t_b)$ as an estimate of $\Omega_2$ and hence $\Delta$ saturates to a relatively large value of around $10^{-10}$.
    \item[Indirect method.] Since this method compares two estimates $C_2$ and $\tilde C_2$ initialized at the same point in time and evolved over the same interval, they converge to each other exponentially fast over a short time interval as they each converge to $\Omega_2$. Over long times, however, the vectors defining these center subspace estimates align/anti-align significantly, resulting in the same numerical issues discussed for the direct method. Since both $C_2$ and $\tilde C_2$ both become increasingly inaccurate estimates of $\Omega_2$ as $t_b$ increases, they have no reason to remain aligned and hence the distance $\Delta$ between the two subspace estimates increases during the backward time evolution.
\end{description}

\begin{figure}[htbp]
    \centering
    \includegraphics[width=0.8\linewidth]{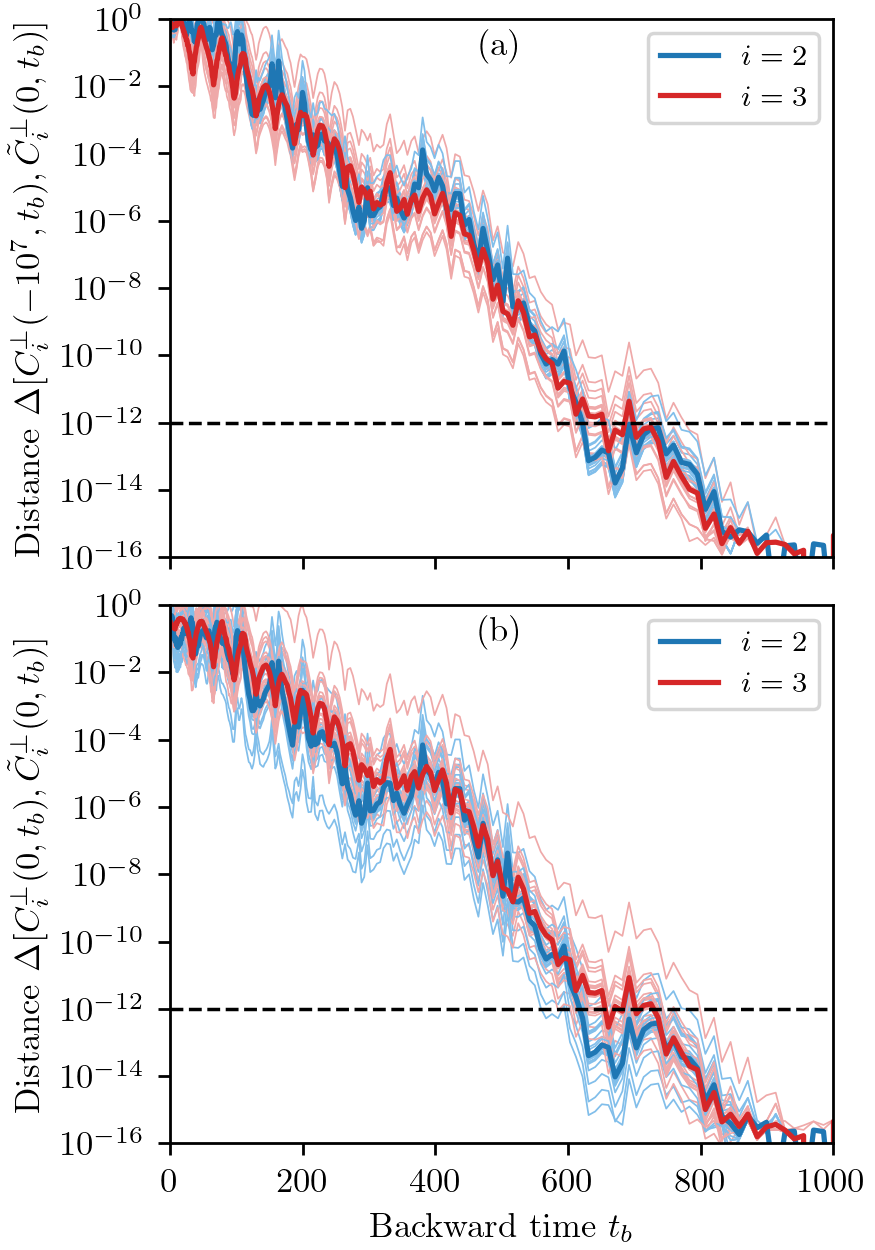}
    \caption{Similar to Fig.~\ref{fig:back_lin}, except the GC algorithm has been used with the center correction, where the CLVs in $C_2$ and $\tilde C_2$ are orthonormalized at each point during backward integration to create $C_2^{\perp}$ and $\tilde C_2^{\perp}$, respectively, while $C_i^{\perp}=C_i$ and $\tilde C_i^{\perp}=\tilde C_i$ for $i\neq2$.}
    \label{fig:back_corr_lin}
\end{figure}

\begin{figure}[htbp]
    \centering
    \includegraphics[width=0.8\linewidth]{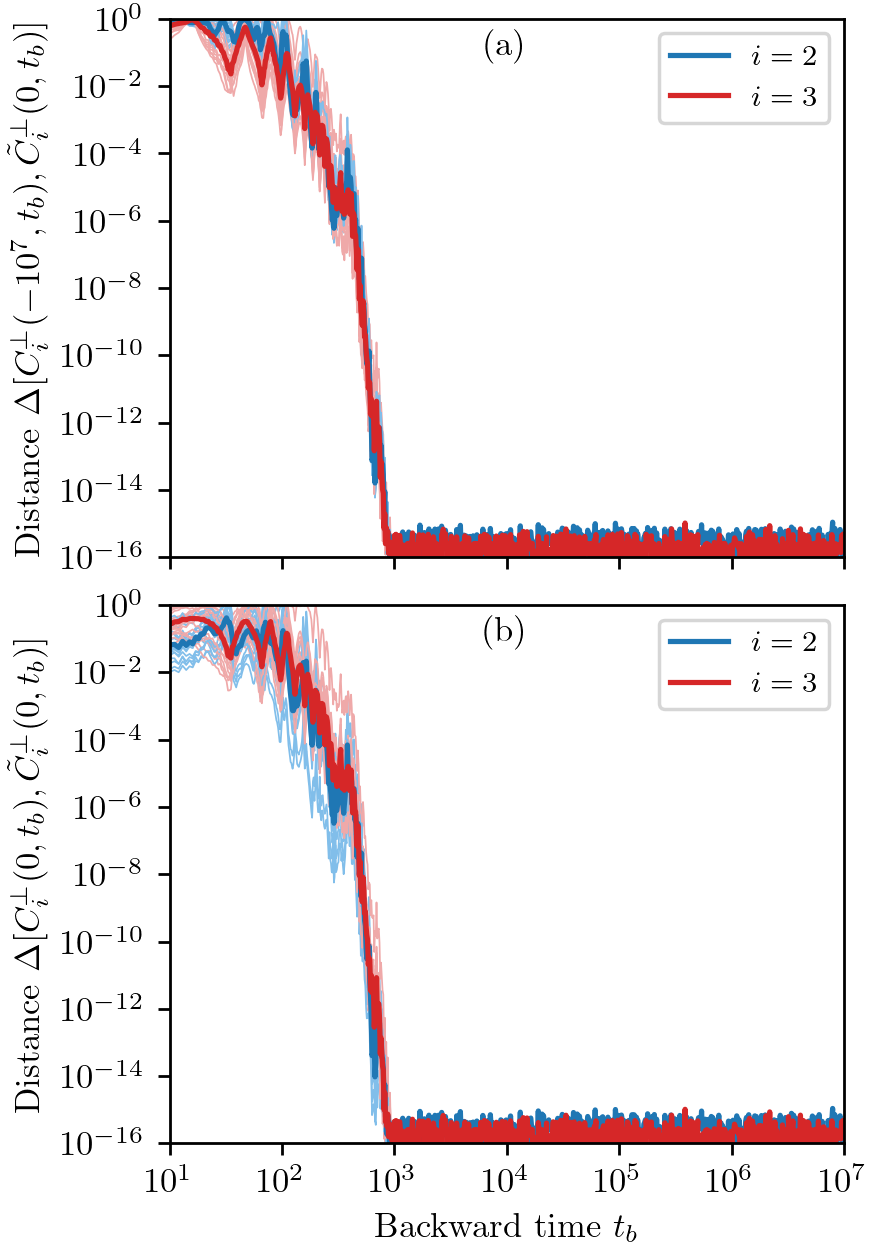}
    \caption{Similar to Fig.~\ref{fig:back_corr_lin}, except the figure is in log-log scale and the backward time evolution of $\Delta$ over the entire backward transient interval of $10^7$ is shown here.}
    \label{fig:back_corr_log}
\end{figure}

Having explained the behavior seen in Fig.~\ref{fig:back_log}, we now propose a simple adaptation of the backward evolution part of the GC algorithm that increases the accuracy of these subspace computations. We propose that, after each step during backward transient and dynamics phases, simply orthonormalize the two CLV estimates used to construct the center subspace estimate (for both $C_2$ and $\tilde C_2$). These orthonormalized CLVs maintain the same span as the original CLVs, but those in the center subspace estimates are prevented from (anti-)aligning, which we have argued is the cause of the poor accuracy of the center subspace estimates $C_2$ and $\tilde C_2$. We refer to this adaptation of the GC algorithm as the \textit{center correction}, and we denote these new estimates of the center subspace by $C_2^{\perp}$ and $\tilde C_2^{\perp}$, respectively. Similarly to Fig.~\ref{fig:back_lin}, in Fig.~\ref{fig:back_corr_lin} we present the backward time evolution of $\Delta$ over the same interval, except the center correction is used for each method. For $i\neq2$, we define $C_i^{\perp}=C_i$ and $\tilde C_i^{\perp}=\tilde C_i$, but for convenience we nonetheless use this new notation for all subspaces when using the center correction. While for small $t_b$ Figs.~\ref{fig:back_lin} and \ref{fig:back_corr_lin} look nearly identical, from around $t_b=600$ in Fig.~\ref{fig:back_corr_lin}(a) we see for the $i=2$ case that $\Delta$ now reliably crosses the $10^{-12}$ threshold for all simulations. Similarly to Fig.~\ref{fig:back_log}, we show in Fig.~\ref{fig:back_corr_log} the backward time evolution of $\Delta$ in log-log scale over a $10^7$ interval when using the center correction; here we see that, regardless of whether the direct or indirect method is used, $\Delta$ decays to a level near machine precision and remains there, even for large $t_b$. This demonstrates the effectiveness of the center correction in improving the accuracy of the center subspace computed by the GC algorithm. Note, however, that we have only tested the applicability of the center correction for 2-D center subspaces, which are common for chaotic orbits of autonomous Hamiltonian systems.

When applying the center correction to the GC algorithm to improve the accuracy of the center subspace computations, we see from Fig.~\ref{fig:back_corr_lin} that for all simulations and for both methods, the distance $\Delta$ between relevant pairs of subspaces decreases below the threshold of $10^{-12}$ by some backward time $t_b\in[580,800]$. Similarly to the forward transient phase, we recommend that the backward transient phase of the GC algorithm with the center correction be stopped as soon as the distance $\Delta$ between computed subspaces $C_i^{\perp}$ and $\tilde C_i^{\perp}$ has decreased below the threshold for every $i=1,2,3$ for that particular simulation, thus ensuring the accuracy of the computed splitting subspaces while avoiding an unnecessarily long transient phase. Again, we note that there is no significant difference between our results for the direct and indirect methods, so we recommend the simpler and faster indirect method.

\subsection{Results for the three degrees of freedom Hamiltonian system}

We now extend our numerical investigation of the H\'enon-Heiles system (see Sect.~\ref{sec:hh_results}) to the three degree of freedom system \eqref{eq:3d_hamiltonian}, yielding analogous results. For our computations, we use a chaotic orbit with initial condition
\begin{align}
    (q_i,p_i)=\left(0,\sqrt{0.06/\omega_i}\right)\label{eq:3d_ic}
\end{align}
for $i=1,2,3$. The first two LEs for this orbit were computed in \cite{SkokosEtAl2007} as $\chi_1\approx0.03$ and $\chi_2\approx0.008$, from which the rest of the LEs are easily inferred due to the LE symmetries for Hamiltonian systems (see Sect.~\ref{sec:clv}). In particular, the only degeneracy in this LE spectrum is the middle pair $\chi_3=\chi_4=0$, so the discussion in Sect.~\ref{sec:hh3D} regarding the various subspace definitions applies here.

\subsubsection{Forward transient phase}
\label{sec:3d_forward}

\begin{figure}[htbp]
    \centering
    \includegraphics[width=0.8\linewidth]{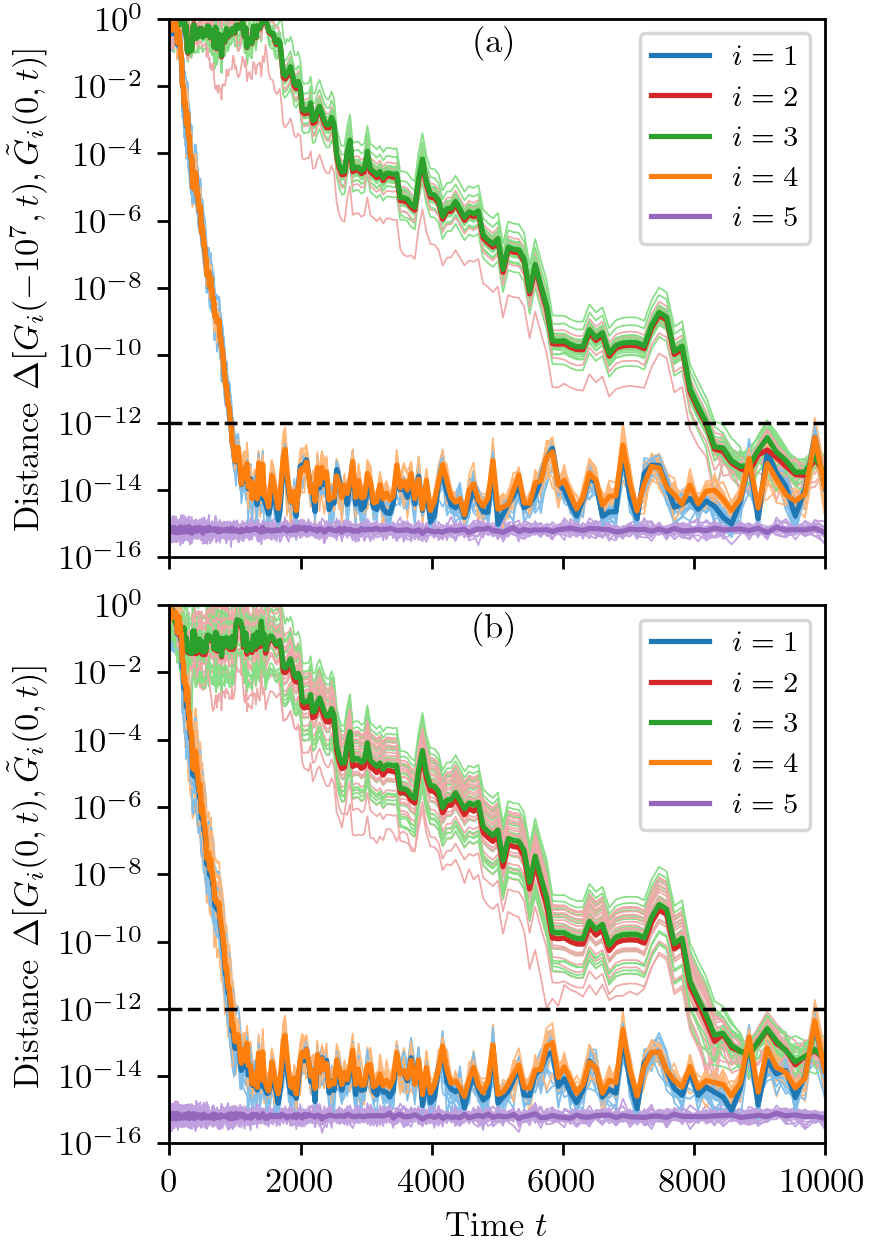}
    \caption{Similar to Fig.~\ref{fig:forward}, but for system \eqref{eq:3d_hamiltonian} with initial condition \eqref{eq:3d_ic} over a forward transient phase of length $10^4$. Note that the thick blue and orange curves practically overlap each other, and the red and green curves similarly overlap.}
    \label{fig:3d_forward}
\end{figure}

Similarly to Sect.~\ref{sec:forward}, we compute the time evolution of the distance $\Delta$ between independent estimates $G_i$ and $\tilde G_i$ of $\Gamma_i^-$ for $i=1,\dots,5$ during the forward transient phase using the direct and indirect methods, which we present in Fig.~\ref{fig:3d_forward} over a time interval of $T=10^4$ time units for 20 simulations that each use a different set of random initial deviation vectors. For the direct method, we again use $T_{\infty}=10^7$. We see from the figure that for $i=1,4$ the threshold $\Delta=10^{-12}$ is first reached for each simulation at some time $t\in[800,1100]$, while for $i=2,3$ it is reached at $t\in[7800,8500]$, regardless of the method used. As soon as these subspaces have converged for every $i=1,\dots,5$, we recommend stopping the forward transient phase of the GC algorithm, which would be at $t\approx8000$ for our simulations.  Note $\Delta$ for $i=5$ is practically zero over the entire time interval since $G_5$ and $\tilde G_5$ coincide with the tangent space. As there is once again no significant difference between our results for the direct and indirect methods, we reaffirm our earlier recommendation of using the indirect method for measuring convergence during the forward transient phase of the GC algorithm, terminating the phase when the threshold is crossed.

\subsubsection{Backward transient phase}

Following the same methodology as in Sect.~\ref{sec:back}, once the forward transient phase for system \eqref{eq:3d_hamiltonian} is complete we evolve the system further forward in time by approximately $10^7$ time units for the forward dynamics phase of the GC algorithm. From here, we begin the backward transient phase and again use a very long backward time interval of $T=10^7$  for this phase to expose the poor center subspace convergence and demonstrate the effectiveness of the center correction. We again use $T_{\infty}=10^7$ for the direct method.

\begin{figure}[p]
    \centering
    \includegraphics[width=0.8\linewidth]{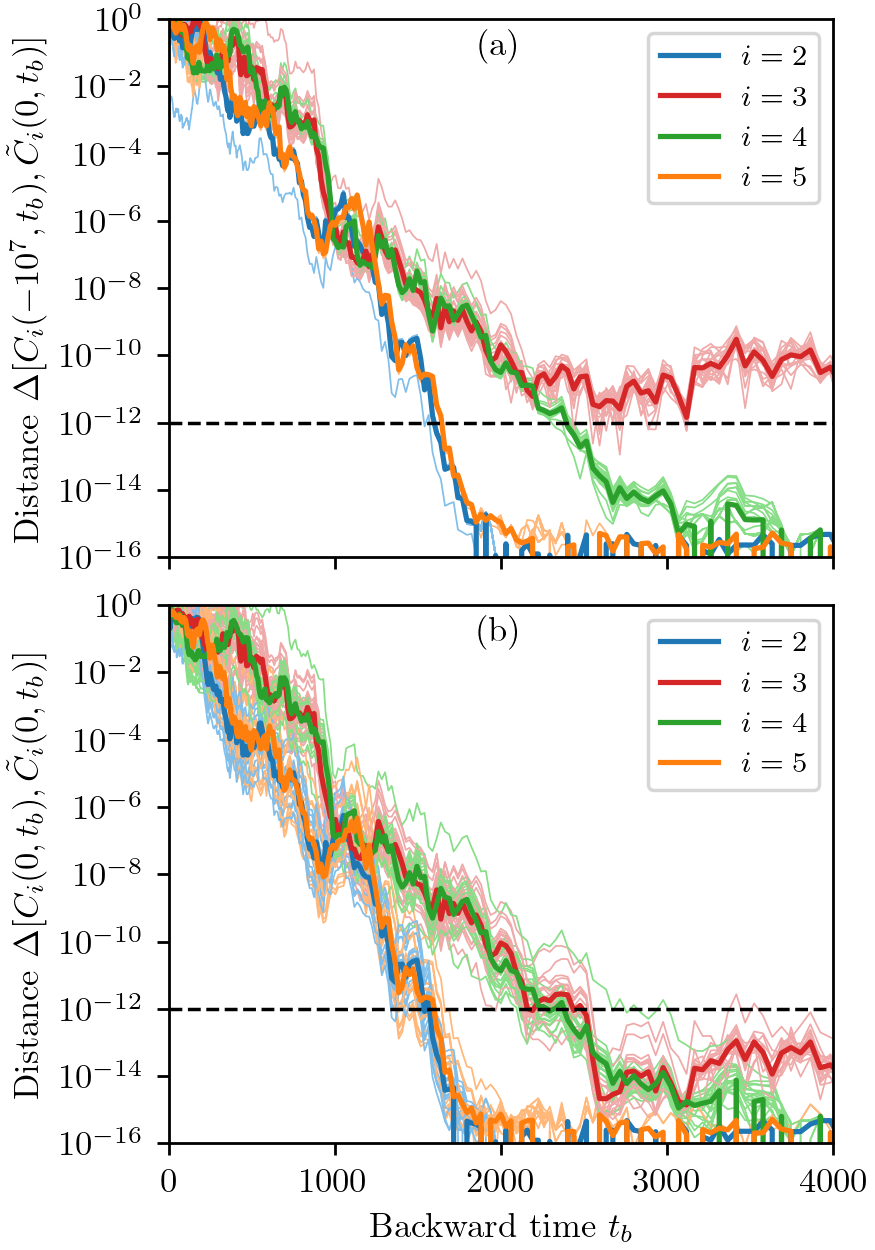}
    \caption{Similar to Fig.~\ref{fig:back_lin}, but for system \eqref{eq:3d_hamiltonian}: the backward time evolution of the distance $\Delta$ computed using the (a) direct and (b) indirect methods during the first $4000$ time units of the backward transient phase of the GC algorithm.}
    \label{fig:3d_back_lin}
\end{figure}

\begin{figure}[p]
    \centering
    \includegraphics[width=0.8\linewidth]{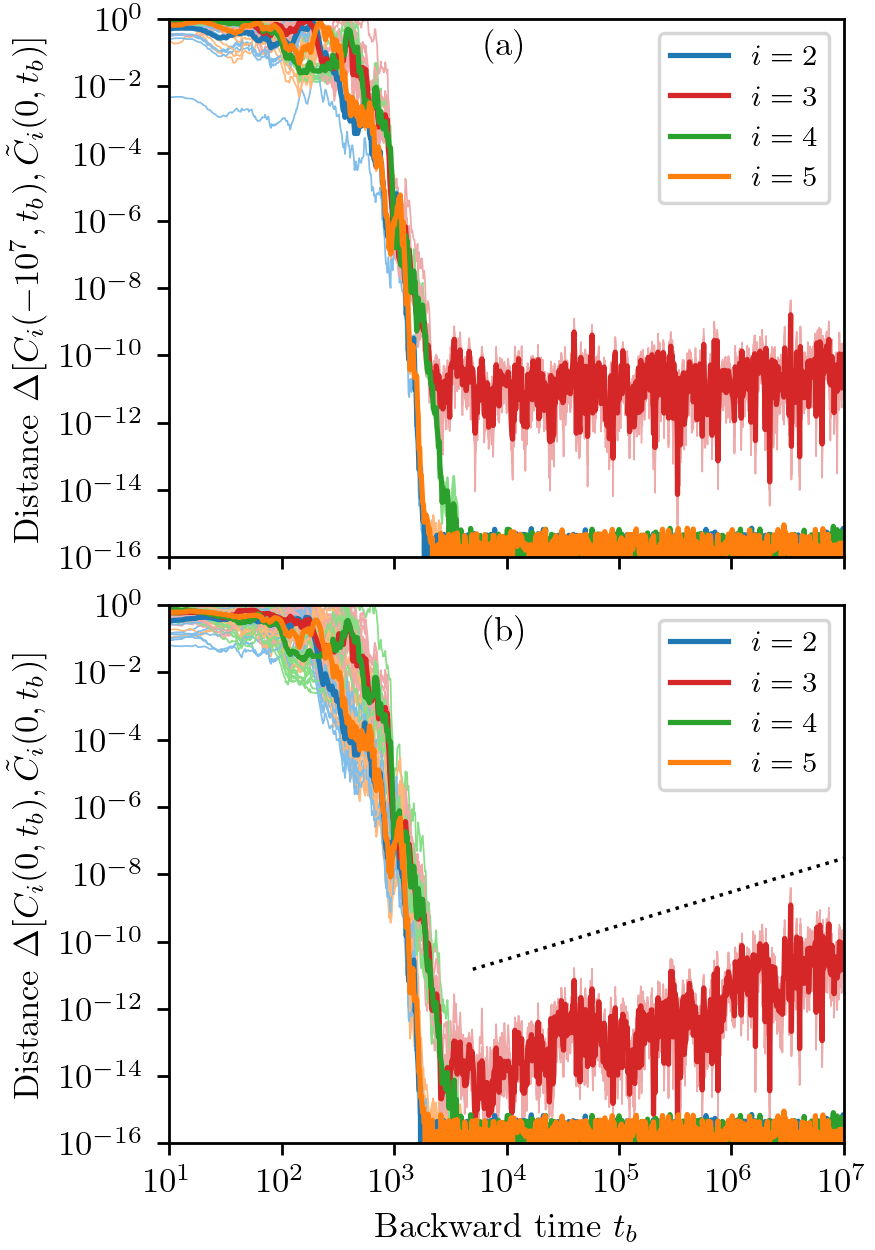}
    \caption{Similar to Fig.~\ref{fig:3d_back_lin}, where panels (a) and (b) correspond between the figures. This figure, however, is in log-log scale and the backward time evolution of $\Delta$ over the entire backward transient interval of $10^7$ is shown here. The black dotted line in (b) denotes a function $\propto t_b$.}
    \label{fig:3d_back_log}
\end{figure}

\begin{figure}[p]
    \centering
    \includegraphics[width=0.8\linewidth]{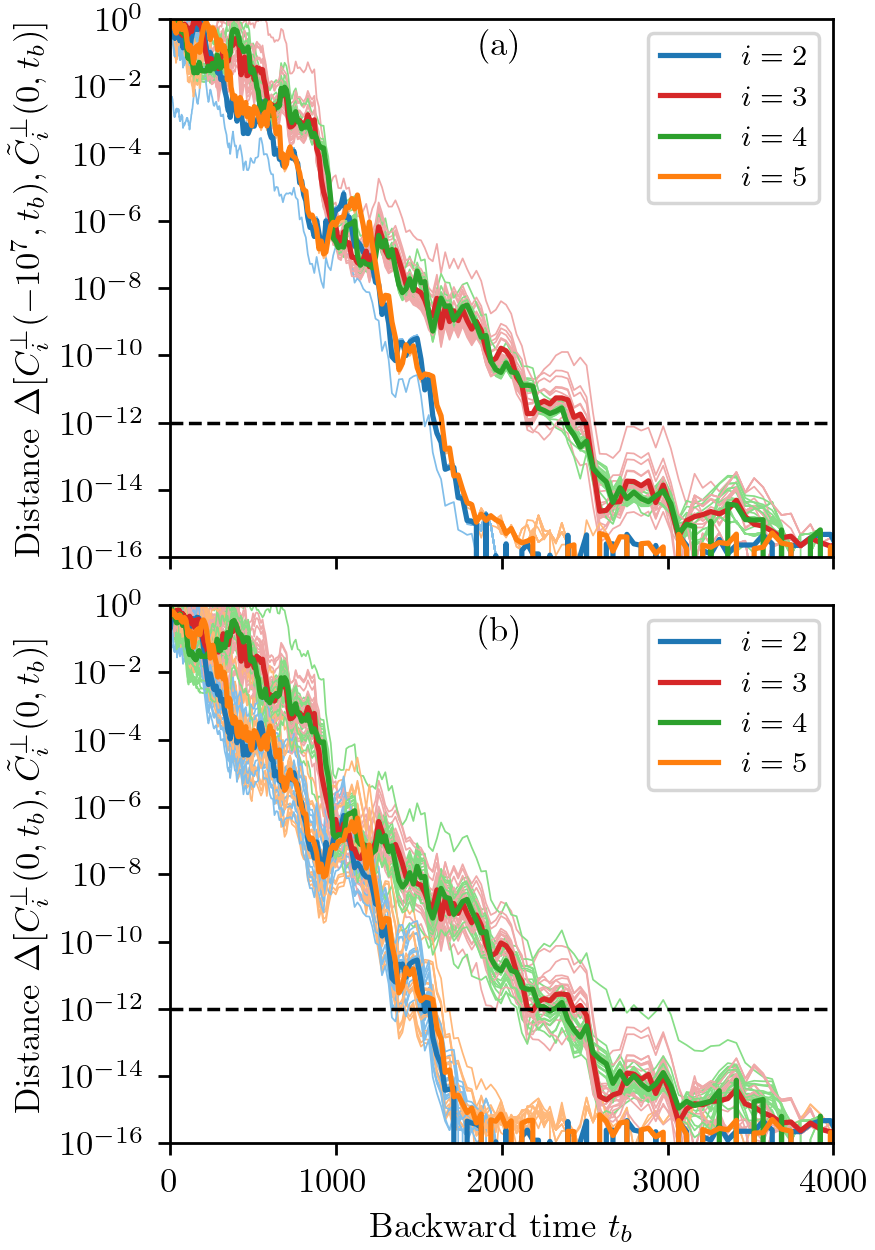}
    \caption{Similar to Fig.~\ref{fig:3d_back_lin}, except the GC algorithm has been used with the center correction, where the CLVs in $C_3$ and $\tilde C_3$ are orthonormalized at each point during backward integration to create $C_3^{\perp}$ and $\tilde C_3^{\perp}$, respectively, while $C_i^{\perp}=C_i$ and $\tilde C_i^{\perp}=\tilde C_i$ for $i\neq3$.}
    \label{fig:3d_back_corr_lin}
\end{figure}

\begin{figure}[p]
    \centering
    \includegraphics[width=0.8\linewidth]{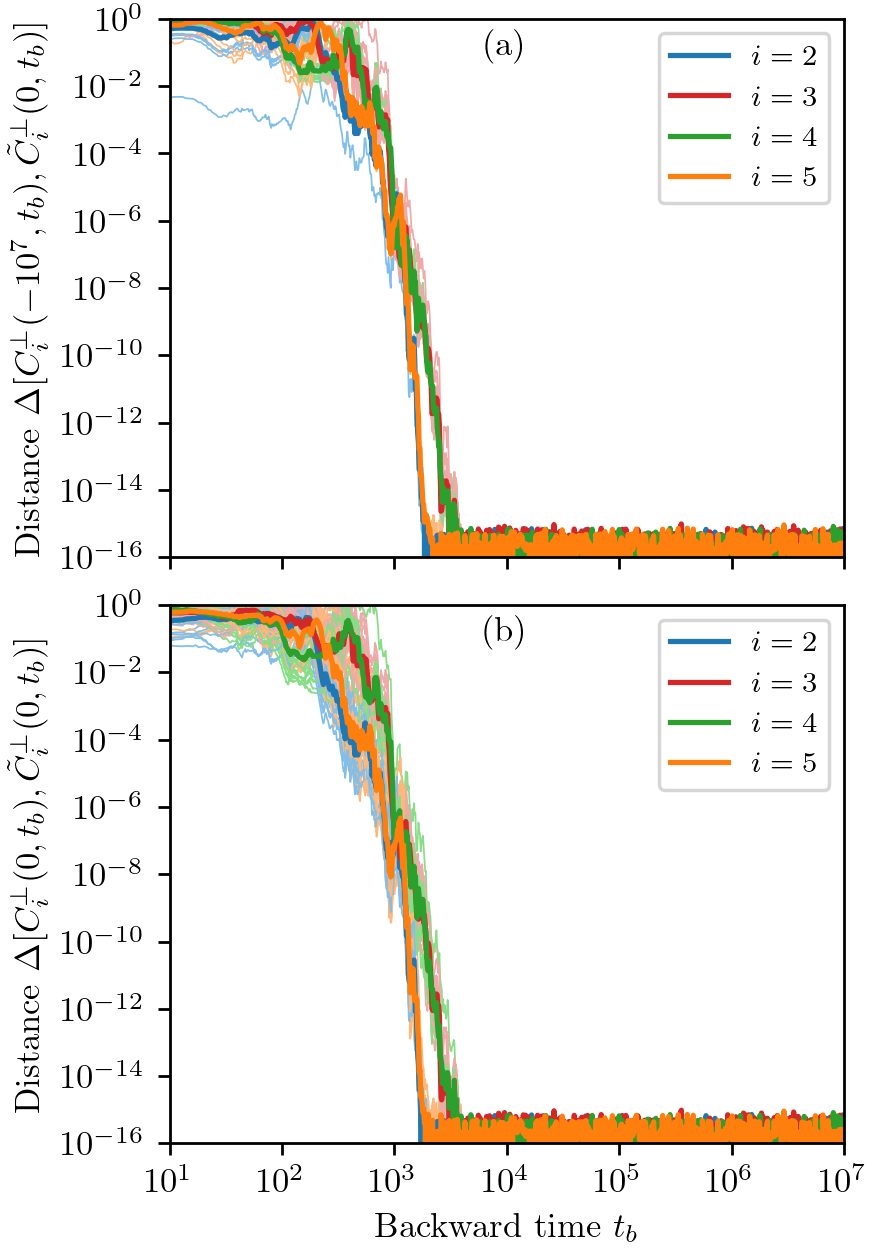}
    \caption{Similar to Fig.~\ref{fig:3d_back_corr_lin}, except the figure is in log-log scale and the backward time evolution of $\Delta$ over the entire backward transient interval of $10^7$ is shown here.\\[1.9em]}
    \label{fig:3d_back_corr_log}
\end{figure}

The backward time evolution of the distance $\Delta$ between estimates $C_i$ and $\tilde C_i$ of $\Omega_i$ for $i=2,3,4,5$ is shown in Figs.~\ref{fig:3d_back_lin} and \ref{fig:3d_back_log} in log-linear and log-log scales, respectively. We see from these figures that the 2-D center subspace estimates $C_3$ and $\tilde C_3$ fail to reliably converge (up to the $\Delta=10^{-12}$ threshold) over long times for both the direct and indirect methods. The reason we propose for this poor convergence is the same as the reason given in Sect.~\ref{sec:back} for the H\'enon-Heiles system: the two CLV estimates in each of the center subspace estimates align/anti-align during backward evolution, resulting in increasingly poor numerical estimates of the center subspace. This issue is once again fixed when using the center correction, the results for which are shown in Figs.~\ref{fig:3d_back_corr_lin} and \ref{fig:3d_back_corr_log}, where we see that the center subspace estimates converge reliably for all simulations over the entire time interval computed, and we see no significant difference between our results for the direct and indirect methods. These results for system \eqref{eq:3d_hamiltonian} provide some evidence that the indirect method for measuring convergence and the center correction adaptation of the GC algorithm apply more generally to autonomous Hamiltonian systems.

\section{Summary and conclusions}\label{sec:conclusions}

After briefly reviewing the relevant theory of CLVs and discussing their computation via the GC algorithm, we used the distance $\Delta$ \eqref{eq:distance} to measure the convergence between relevant subspaces in both the H\'enon-Heiles system \eqref{eq:hamiltonian} and a Hamiltonian system with three degrees of freedom \eqref{eq:3d_hamiltonian}. We proposed two methods, a direct one and an indirect one, for determining the level of convergence between relevant subspaces during the transient phases of the GC algorithm, and we found the time evolution of $\Delta$ to be very similar for both methods during the forward transient phase. For the backward transient phase, however, we noticed that the accuracy of the computed 2-D center subspace is poor, particularly when computed over long times intervals. By proposing a small adaptation of the GC algorithm, which we call the center correction, we improved the accuracy and stability of the algorithm when used to compute this subspace. With the center correction, we found that the two methods again produced practically the same results. As a matter of pragmatism, we recommended the indirect method as an efficient means of checking the convergence of the relevant subspaces computed using the GC algorithm since it requires less CPU time to compute than the direct method, which requires a costly pre-computation.

The main advisory outcomes of this work are therefore summarized as follows:

\begin{description}[style=nextline,font=\small\bfseries]
    \item[When to stop the transient phases of the GC algorithm]
    When computing the forward transient phase of the GC algorithm, initialize two sets of deviation vectors which define the subspaces $G_i$ and $\tilde G_i$ for $i=1,\dots,m$, where $m$ is the number of distinct LEs in the spectrum \eqref{eq:spec}. Evolve both sets of deviation vectors independently according to the GC algorithm while frequently computing the distance $\Delta$ between each corresponding pair of subspaces $G_i$ and $\tilde G_i$. As soon as $\Delta$ decreases below some small threshold (e.g.\ $10^{-12}$) for every pair of subspaces, then the subspaces $G_i$ and $\tilde G_i$ have converged to each other, so we can assume that $G_i$ has converged to $\Gamma_i^-$ and thus stop the forward transient phase. The backward transient phase is completely analogous: initialize two sets of deviation vectors $C_i$ and $\tilde C_i$, evolve both sets of vectors independently according to the GC algorithm, then stop the backward transient phase once $\Delta[C_i,\tilde C_i]$ decreases below the chosen threshold for all $i=1,\dots,m$.
    \item[How to accurately compute the 2-D center subspace]
    In order to accurately compute the 2-D center subspace over long intervals of time, we advise using the center correction adaptation of the GC algorithm: during any backward integration (including both the backward transient and backward dynamics phases), regularly orthonormalize the two middle deviation vectors which lie in the center subspace estimate to prevent their (anti-)alignment and thus maintain the accuracy of the estimated 2-D center subspace which they span.
\end{description}

We hope that these proposed augmentations to the GC algorithm will assist researchers in their CLV computations by determining when to end the transient phases and improving the accuracy of their center subspace computations. Although we have presented numerical results exclusively for autonomous Hamiltonian systems, there is no apparent reason why the proposed methods for terminating the transient phases of the GC algorithm could not be extended to more general settings, such as dissipative systems. Verifying this extension would constitute a valuable direction for future work.


\section*{Acknowledgments}
J.-J.~d.~P.\ was supported by the David and Elaine Potter Fellowship, the VC Research Scholarship of the University of Cape Town (UCT), and the UCT Master's Research Scholarship. M.~H. was supported by the Max Planck Society and the National Research Foundation (NRF) of South Africa, Grant No.\ 129630. The authors thank the center for High Performance Computing of South Africa (CHPC) for providing the computational resources used in this work.

\appendix
\section{A toy model for understanding CLVs}
\label{sec:appendix}
In order to supplement the necessarily technical discussion on CLVs and the related subspaces given in Sect.~\ref{sec:clv}, we provide here an example of a simple dynamical system with a tangent space of $\mathbb R^3$ wherein these quantities can be easily understood and visualized. Note that this example is not a Hamiltonian system, so we only explore it here for pedagogical purposes.

Consider the system with equations of motion $\dot x_i=\lambda_i x_i$, where $i=1,2,3$ and $\lambda_1>\lambda_2>\lambda_3>0$. The variational equations are thus $\dot w_i=\lambda_i w_i$, which have solutions $w_i(t)=w_i(0)\exp(\lambda_i t)$, and thus the LEs are simply $\lambda_i$. For a generic deviation vector $\bs w(t)=\sum_{i=1}^3w_i(t)\hat{\bs e}_i$, the $\hat{\bs e}_1$ component dominates the others for large $t$, so $\left\lVert\bs w(t)\right\lVert\sim\exp(\lambda_1 t)$ as $t\to\infty$ when $w_1$ is non-zero. However, in the special case where $w_1=0$ and $w_2\ne0$, this exponential growth rate would instead be $\lambda_2$, and if only $w_3$ is non-zero, then that rate would be $\lambda_3$. By making the following subspace definitions, we can formalize the classification of deviation vectors by their asymptotic growth rates: $\Gamma_1^+=\spn(\hat{\bs e}_1,\hat{\bs e}_2,\hat{\bs e}_3)$, $\Gamma_2^+=\spn(\hat{\bs e}_2,\hat{\bs e}_3)$, $\Gamma_3^+=\spn(\hat{\bs e}_3)$, and $\Gamma_4^+=\{0\}$. From these subspace definitions, which form a filtration \eqref{eq:filter4} of the tangent space, it follows that if $\bs w\in\Gamma_i^+\setminus\Gamma_{i+1}^+$ then $\lambda^+(\bs w)=\lambda_i$, where $i=1,2,3$ and $\lambda^+(\bs w)$ is the exponential growth rate of $\left\lVert\bs w(t)\right\lVert$ as $t\to\infty$, see \eqref{eq:growth}.

Considering now the tangent dynamics as $t\to-\infty$, the asymptotic exponential growth rates $\lambda_i$ of the deviation vector components $w_i(t)$ become decay rates under time reversal. Therefore, $\left\lVert\bs w(t)\right\lVert\sim\exp(\lambda_3 t)$ as $t\to-\infty$ if $w_3\ne0$ since the $\hat{\bs e}_3$ component of $\bs w(t)$ dominates the others, but if $w_3=0$ and $w_2\ne0$ then the rate of decay is governed by $\lambda_2$, and if $w_1$ is the only non-zero component then the rate is $\lambda_1$. Formally: if $\bs w\in\Gamma_i^-\setminus\Gamma_{i-1}^-$ then $\lambda^-(\bs w)=\lambda_i$, where $\lambda^-$ is given in \eqref{eq:growth} and we define the subspaces $\Gamma^-_i$ in this case as $\Gamma^-_0=\{0\}$, $\Gamma^-_1=\spn(\hat{\bs e}_1)$, $\Gamma^-_2=\spn(\hat{\bs e}_1,\hat{\bs e}_2)$, and $\Gamma^-_3=\spn(\hat{\bs e}_1,\hat{\bs e}_2,\hat{\bs e}_3)$, which forms another filtration \eqref{eq:filter4} of the tangent space.

Finally, a splitting \eqref{eq:splitting} of the tangent space is constructed by intersecting pairs of filtration subspaces: $\Omega_i = \Gamma_i^+ \cap \Gamma_i^-$. In our example, $\Omega_1=\spn(\hat{\bs e}_1)$, $\Omega_2=\spn(\hat{\bs e}_2)$, and $\Omega_3=\spn(\hat{\bs e}_3)$. Under the evolution of the tangent dynamics, any deviation vector in $\Omega_i$ will grow exponentially at a rate of $\lambda_i$ as $t\to\infty$, while as $t\to-\infty$ it will decay at an exponential rate of the same LE $\lambda_i$. The CLVs are any set of $N$ linearly independent covariant unit vectors from the splitting subspaces $\Omega_i$. In this case, since each $\Omega_i$ is 1-D, we can uniquely (up to signs) choose our three CLVs as follows: $\hat{\bs e}_1$, $\hat{\bs e}_2$, and $\hat{\bs e}_3$.

\section{Principal angles}
\label{sec:appendix2}
Since our definition of distance $\Delta$ between linear subspaces given in \eqref{eq:distance} depends on the notion of principal angles, we give here a brief overview of the subject. Following the definitions given in \cite{BjorckGolub1973,GolubVanLoan1996}, let $U$ and $V$ be linear subspaces of some real inner product space that satisfy
\begin{align}
    \dim U \ge \dim V = q \ge 1.
\end{align}
The principal angles $\theta_1,\dots,\theta_q\in[0,\pi/2]$ between $U$ and $V$ are defined recursively for $k=1,\dots,q$ by
\begin{align}
    \cos(\theta_k) = \max_{\hat{\bs u}\in U}\, \max_{\hat{\bs v}\in V}\, (\hat{\bs u}\cdot\hat{\bs v}) = \hat{\bs u}_k\cdot\hat{\bs v}_k
\end{align}
subject to
\begin{align}
    \begin{split}
        &\left\lVert\hat{\bs u}\right\lVert = \left\lVert\hat{\bs v}\right\lVert=1,\\
        &\hat{\bs u}\cdot\hat{\bs u}_i = \hat{\bs v}\cdot\hat{\bs v}_i = 0,\quad i=1,\dots,k-1,
    \end{split}
\end{align}
where the unit vectors $\hat{\bs u}_i$ and $\hat{\bs v}_i$ are known as the principal vectors between $U$ and $V$. Note that while the principal vectors are not uniquely defined, the principal angles are. The principal angles satisfy $\theta_1\le\cdots\le\theta_q$, where the smallest principal angle $\theta_1$ has a simple interpretation: it is the minimum/smallest angle between any pair of non-zero vectors $\bs u\in U$ and $\bs v\in V$. The largest principal angle $\theta_\text{max}:=\theta_q$ is zero if and only if $U=V$, thus motivating its role in our measure of distance $\Delta$ between two equidimensional subspaces defined in \eqref{eq:distance}.

\begin{figure}[htbp]
    \centering
    \includegraphics[width=\linewidth]{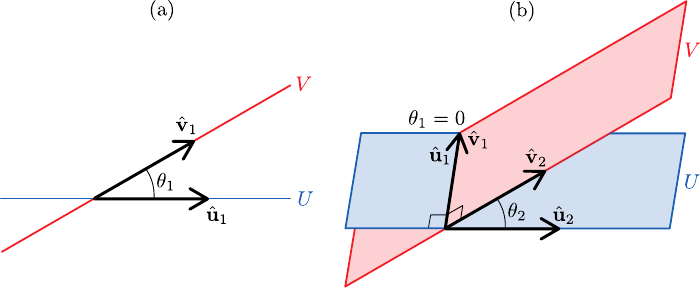}
    \caption{An illustration of the principal angles and corresponding vectors (black) between the subspaces $U$ (blue) and $V$ (red) in $\mathbb R^3$, for the cases where both subspaces are (a) 1-D and (b) 2-D.}
    \label{fig:angles}
\end{figure}

We give two simple  examples here where $U\neq V$ are equidimensional subspaces of $\mathbb R^3$. If the subspaces are 1-$D$, then there is only one principal angle between $U$ and $V$, as depicted in Fig.~\ref{fig:angles}(a). If the subspaces are 2-$D$, as shown in Fig.~\ref{fig:angles}(b), then there are two principal angles: $\theta_1=0$, which vanishes due to the non-trivial intersection between $U$ and $V$, and $\theta_2>0$, which does not vanish because $U\neq V$.

Various algorithms exist for the computation of principal angles. Two such algorithms using singular value decomposition were proposed in \cite{BjorckGolub1973}. However, when computing some principal angle $\theta_k$, both algorithms suffer a significant loss of accuracy when $\theta_k$ is near a certain value: for the one algorithm, this occurs when $\theta_k\approx0$, while for the other, it occurs when $\theta_k\approx\pi/2$. These two algorithms were later combined into a hybrid algorithm \cite{KnyazevArgentati2002} that avoids this issue by ensuring that the more accurate of the two computations is used for all values of $\theta_k\in[0,\pi/2]$. We use this hybrid algorithm, as implemented in the SciPy package \cite{VirtanenEtAl2020}, for all our computations in this paper involving principal angles.

\bibliography{references}

\end{document}